\documentclass[journal,transmag]{IEEEtran}
\ifCLASSINFOpdf
   \usepackage[pdftex]{graphicx}
\else
\fi
\begin{document}
%
\title{Network Growth From Global and Local Influential Nodes}



\author{Jiaojiao~Jiang,
		and~Sanjay~Jha
\IEEEcompsocitemizethanks{\IEEEcompsocthanksitem J. Jiang and S. Jha are with the School of Computer Science and Engineering, University of New South Wales, Sydney, NSW 2032, Australia. Email: \{jiaojiao.jiang, sanjay.jha\}@unsw.edu.au}
}

%



\IEEEtitleabstractindextext{%
\begin{abstract}
In graph theory and network analysis, node degree is defined as a simple but powerful centrality to measure the local influence of node in a complex network. Preferential attachment based on node degree has been widely adopted for modeling network growth. However, many evidences exist which show deviation of real network growth from what a pure degree-based model suggests. It seems that node degree is not a reliable measure for predicting the preference of newcomers in attaching to the network, or at least, it does not tell the whole story. In this paper, we argue that there is another dimension to network growth, one that we call node ``coreness''. The new dimension gives insights on the global influence of nodes, in comparison to the local view the degree metric provides. We found that the probability of existing nodes attracting new nodes generally follows an \textit{exponential} dependence on node coreness, while at the same time, follows a \textit{power-law} dependence on node degree. That is to say, high-coreness nodes are more \textit{powerful} than high-degree nodes in attracting newcomers. The new dimension further discloses some hidden phenomena which happen in the process of network growth. The power of node degree in attracting newcomers increases over time while the influence of coreness decreases, and finally, they reach a state of \textit{equilibrium} in the growth. All these theories have been tested on real-world networks. 
\end{abstract}

\begin{IEEEkeywords}
Preferential attachment, coreness, degree.
\end{IEEEkeywords}}

\maketitle

\IEEEdisplaynontitleabstractindextext

%
\IEEEpeerreviewmaketitle

\section{Introduction}

Current studies on preferential attachment mainly focus on the local influence of nodes, such as degree \cite{NM2001}, triadic closures \cite{KP2013}, community \cite{WT2013}, and so forth.
To date, one of the most popular mechanisms to model growing networks remains degree-based preferential attachment \cite{NM2001,SG2016,nsour2020hot}. It is hypothesized that the probability $T(k)$ of a node of degree $k$ attracting new nodes is a monotonically increasing function of $k$, \textit{i.e.}, $T(k)\propto k^{\alpha}$. In many real network systems, it is verified that, the preferential attachment rate $\kappa(k)=\int_0^kT(k')dk'$ shows a \textit{power-law} dependence on $k$, \textit{i.e.}, $\kappa(k)\propto k^{\alpha+1}$. However, many evidences exist and show deviation of real network growth from what the solely degree-based model suggests. For example, it fails to explain the dense modular structures \cite{NM2002}, high clustering \cite{WD1998}, and its most significant feature$-$the scale-free degree distribution$-$collapses in equilibrium situations \cite{MC2006,BJ2013}. Later, the interaction between the neighborhood \cite{BK2004,JJ2002} instead of the pure degree of nodes were later taken into account in preferential attachment models. For example, Bagrow and Brockmann \cite{BJ2013} investigated network growth based on the triadic closures of nodes, which quantifies how close the neighbors of a node are to being a clique. Li et. al \cite{LM2014} explored the patterns of link creation in network evolution based on neighborhood assortativity, which characterizes the degree relation between neighboring nodes. The preferential attachment model was further extended by considering the local communities of nodes \cite{GM2007,WT2013,ZK2015,PP2006}. In essence, all these models only consider the local view of nodes. The variousness of the models for preferential attachment also have shed light on the unreliability of the local view of nodes in modeling growing networks, prompting us to wonder whether the global view of nodes is more reliable and stable in network growth.

In this paper, we study the preferential attachment from a new dimension, node ``coreness'' \cite{KM2010}, which measures the global influence of nodes. Intuitively, the global influence of nodes is more stable, which generally dominates the evolution process of a network. To mathematically test this conjecture, we first develop a method to mathematically measure the preferential attachment rate $\kappa(c)$ with respect to node coreness $c$. We find that the probability of existing nodes attracting new nodes generally follows an \textit{exponential} dependence on node coreness, $\kappa(c)\propto e^{\beta c}$, while at the same time, follows a \textit{power-law} dependence on node degree, $\kappa(k)\propto k^{\alpha+1}$. From a mathematical point of view, the increasing rate of exponential functions is significantly faster than that of power-law functions. Hence, coreness does show the power of dominance in network growth. To investigate some hidden phenomena happening in the process of network growth, we compare the preferential attachment rates $\kappa(c)$ and $\kappa(k)$ over time. We find that the power of node degree in attracting newcomers increases over time while the influence of coreness decreases, and finally, they reach a state of equilibrium. To further measure both the local and global influence of nodes in network growth, we develop a measure which can quantify the hybrid preferential attachment probability $T(c,k)$ with respect to both $c$ and $k$ of nodes. We derive a general form for the probability $T(c,k)\propto e^{\beta c+\alpha\mbox{ln}(k)}$. Based on $T(c,k)$, we can evaluate the probability $T(c_0,k)$ of local preferential attachment for nodes of the same coreness $c_0$, and the probability $T(c,k_0)$ for the nodes of the same degree $k_0$. Based on extensive analysis on real temporal networks, we summarize our findings as follows:
\begin{itemize}
  \item We develop a method to measure the preferential attachment rate based on node \textit{coreness} $c$, instead of the pure node \textit{degree}. We find that the probability of existing nodes attracting new nodes shows an \textit{exponential} dependence on $c$, which demonstrates the dominance of node coreness in growing networks. 
  \item We analyze the interplay of degree and coreness in network growth, and find that the power of degree-based preferential attachment \textit{increases} over time and the power of coreness-based preferential attachment \textit{decreases}, and they finally reach to a state of \textit{equilibrium}.
  \item We derive a general form for the probability $T(c,k)$ of \textit{hybrid preferential attachment} with respect to $c$ and $k$. This form not only generates the preferential attachment rates $\kappa(k)$ and $\kappa(c)$, but also guides us in analyzing localized preferential attachment $T(c,k_0)$ and $T(c_0,k)$.
\end{itemize}

The following sections are organized as follows. Section \ref{preliminaries} introduces the preliminaries for this paper. Section \ref{sec:PAcore} presents methods to measure coreness-based preferential attachment and the hybrid preferential attachment. Sections \ref{subsec:PAMCoreness} and \ref{SecHybrid} study these two preferential attachments in temporal networks, respectively. Section \ref{conclusion} presents some concluding remarks.

\section{Preliminaries}\label{preliminaries}

\subsection{Degree-based Preferential Attachment}\label{growth}
During the past two decades, many models have been proposed to capture the topological evolution of complex networks. In particular, a class of models, which are based on the \textit{growth} and \textit{preferential attachment} in networks, is quite successful in reproducing the structural features of real network systems. The \textit{growth} models capture the process that networks continuously expand through the addition of new nodes and links between the nodes, while the \textit{preferential attachment} models learn the rate $T(v)$ with which a node with value $v$ of a local feature (such as degree, clustering coefficient, or assortative coefficient) attracting new links is a monotonically increasing function of $v$. These models include the traditional BA model \cite{BA2000}, distance-dependent preferential attachment (DDPA) evolving model \cite{LM2006} of weighted networks, acquaintance network model \cite{DJ2002} and connecting nearest-neighbor (CNN) model \cite{VA2003}. They have been widely used to simulate the highly interconnected nature of various social, metabolic and communication networks.
\begin{figure}[b]
\centering
\includegraphics[width=0.2\textwidth]{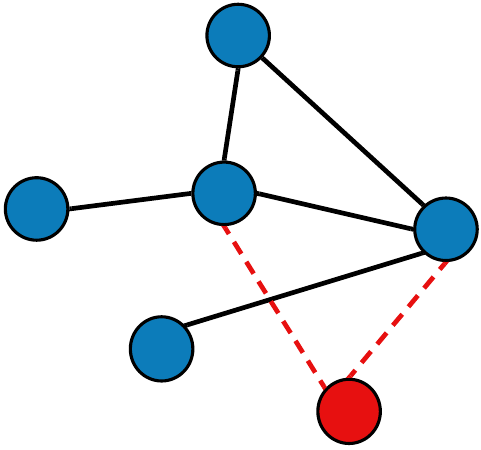}
\caption{\small{The link formation process in the degree-based preferential attachment model.
}}
\label{fig:LinkFormation}
\end{figure}

Here, we introduce a widely used \textit{temporal data} based preferential attachment model \cite{NM2001, JH2003, peng2017social}. The basic idea is to investigate whether new links are likely to attach to nodes with larger degree. In Fig. \ref{fig:LinkFormation}, the degree-based preferential attachment model is illustrated. The new node (red) generates two links connecting to two largest degree nodes. Mathematically, for a newly added node $i$, we use $a_{i\lambda}=1$ to denote that node $i$ attaches to node $\lambda$ within a short time period $\Delta t$, otherwise $a_{i\lambda}=0$. To measure the probability that a new link attaches to a node, which has a degree of $k$ at time $t$, we calculate $n_k(t)$ the number of nodes with degree $k$ at time $t$. We use $N(t)$ to denote the number of nodes at time $t$. Then, the empirical value of the probability $T(k)$ that a new link formed within a short time period $\Delta t$ connects to a node, which has a degree of $k$ at time $t$, can be calculated as follows,
\begin{equation}\label{Tk}
T(k) =\frac{\frac{A(k)}{n_k(t)}}{\displaystyle\sum_{k'}\frac{A(k')}{n_{k'}(t)}},
\end{equation}
where, $A(k)=\displaystyle\sum_{i,\lambda}^{k_{\lambda}(t)=k} a_{i\lambda}$ is the number of nodes with exact degree $k$ at time $t$, but creating new links within next small time period $\Delta t$; and the condition $k_{\lambda}(t)=k$ denotes that node $\lambda$ has degree $k$ at time $t$.

The degree-based preferential attachment \textit{hypothesis} \cite{barabasi1999emergence} states that the rate $T(k)$ with which a node of degree $k$ attracts new links is a monotonically increasing function of degree $k$, namely
\begin{equation}\label{eq3}
T(k)=\frac{k^{\alpha}}{\displaystyle\sum_kk^{\alpha}} \propto k^{\alpha},
\end{equation}
where $\alpha$ is the exponent. In other words, the preferential attachment rate $T(k)$ follows a \textit{power-law} distribution. For example, in the BA model \cite{BA2000}, the power-law exponent $\alpha=1$. 

To obtain a smooth curve from noisy data, the following cumulative function form rather than $T(k)$ is often used:
\begin{equation}\label{eq4}
\kappa(k)=\int_0^kT(k')dk'.
\end{equation}
According to Eqs. (\ref{eq3}) and (\ref{eq4}), $\kappa(k)$ is proportional to $k^{\alpha+1}$, namely
\begin{equation}\label{eq14}
\kappa(k)\propto k^{\alpha+1}.
\end{equation}
This model has been extensively used to measure the preferential attachment phenomenon in different kinds of networks \cite{DB2007, shi2009user, leskovec2008microscopic, BJ2013, LM2014}.


\subsection{$k$-coreness Centrality}\label{kcore}
So far, scientists have proposed a number of measurements and centralities to quantify the influence of nodes in networks, such as PageRank \cite{bianchini2005inside}, clustering coefficient \cite{MA2006}, leverage centrality \cite{joyce2010new}, closeness \cite{NM2010}, betweenness \cite{FL1979}, eigenvector centrality \cite{BP1987}, Katz centrality \cite{KL1953}, degree \cite{AR2000}, and $k$-coreness centrality \cite{KM2010}. 
Some centrality measurements, such as degree centrality and leverage centrality, capture local information (e.g., with respect to immediate nodal neighbours), whereas others, such as betweenness and closeness, quantify how a node is situated within the global network context. The $k$-coreness centrality belongs to the later type of measurements.

\begin{figure}[t]
\centering
\includegraphics[width=0.32\textwidth]{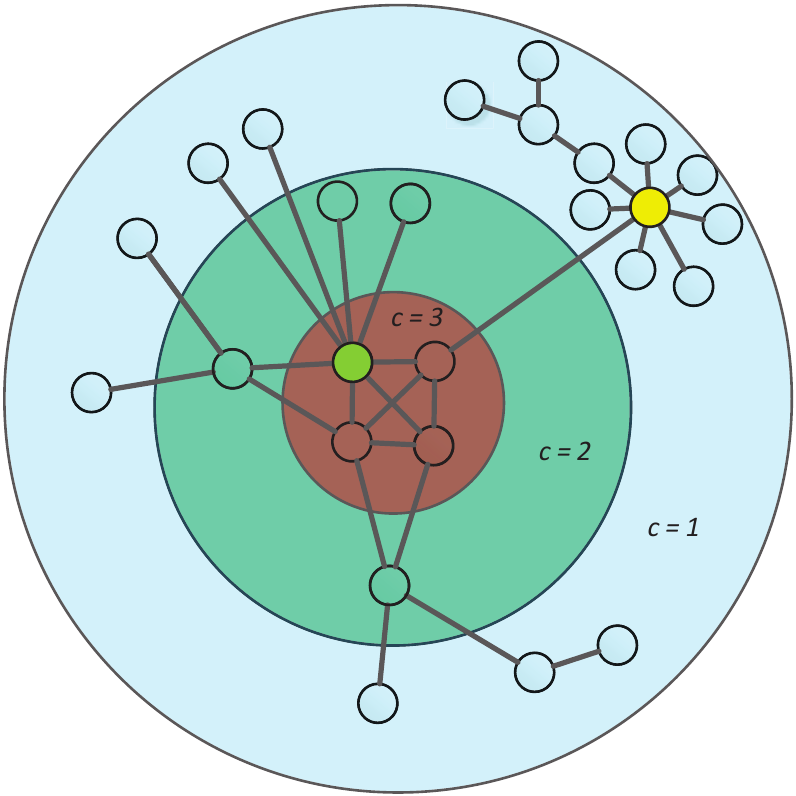}
\caption{\small{ A schematic representation of a network under the $k$-core decomposition. The two nodes of degree $k$ = 8 (green and yellow nodes) are in different locations: one lies at the periphery with $c$ = 1, whereas the other one is in the innermost core with $c$ = 3.
}}
\label{k-core}
\end{figure}

Given a network, the $k$-coreness centrality is gained from the $k$-core structure of the network \cite{KM2010}, to measure the global influence of nodes in the network. The $k$-core structure of a network is obtained by recursively removing all nodes with degree smaller than $k$, until the degree of all remaining nodes is larger than or equal to $k$. This process assigns a \textbf{\textit{coreness}}, denoted by $c$, to each node, representing its location according to successive layers in the network. The greater the coreness $c$, the more influential of the node. Therefore, the most influential nodes have the greatest coreness \cite{KM2010}. The nodes with the same coreness form a \textbf{\textit{shell}} or \textbf{\textit{core}} of a network. Therefore, small values of $c$ define the periphery of the network and large $c$ corresponds to the core of the network. Fig. \ref{k-core} shows an example of the $k$-core structure of a network. We can see that, even though the yellow node and the green node both have 8 neighbors (\textit{i.e.}, with the same value of degree), the green node lies in the innermost core with $c$ = 3, while the yellow node belongs to the periphery with $c$ = 1. Hence, nodes with the same local influence could present different performances on their global influence. In this paper, we will study the dominance of node coreness than node degree in network growth and node preferential attachment process.

\section{Coreness-based Preferential Attachment}\label{sec:PAcore}
In this section, we first develop a method to measure coreness-based preferential attachment during network growth, and then we perform empirical analysis of the model on real-world temporal networks. 
The basic idea is to investigate whether new links are more likely to attach to nodes with higher coreness and study how the local influence (wrt. degree) and the global influences (wrt. coreness) of nodes interplay with each other during network growth. The method is extended from the measurement of degree-based preferential attachment rate $\kappa(k)$ in Section \ref{growth}.

\subsection{Preferential Attachment Model}\label{subsec:PAMCore}
We first calculate $A(c)$, the number of nodes with exact coreness $c$ at time $t$ but creating new links within next small time period $\Delta t$ as follows:
\begin{equation}\label{eq.Ac}
    A(c) = \displaystyle\sum_{i,\lambda}^{c_{\lambda}(t)=c}a_{i\lambda},
\end{equation}
where the condition $c_{\lambda}(t)=c$ denotes that node $\lambda$ has coreness $c$ at time $t$.
To measure the probability that a new link attaches to a node, which has a coreness of $c$ at time $t$, we calculate $n_c(t)$ the number of nodes with coreness $c$ at time $t$. We use $T(c)$ to denote the probability that a new link formed within a short time period $\Delta t$ connects to a node, which has a coreness of $c$ at time $t$.
Then, similarly to Eq. (\ref{Tk}), the empirical value of the probability $T(c)$ can be calculated as follows:
\begin{equation}
T(c) = \frac{\frac{A(c)}{n_c(t)}}{\displaystyle\sum_{c'}\frac{A(c')}{n_{c'}(t)}},
\end{equation}
where $A(c)$ is defined in Eq. (\ref{eq.Ac}).

In degree-based preferential attachment, the \textit{hypothesis} states that the rate $T(k)$ with which a node with degree $k$ attracts new links is a \textit{power-law} function of $k$, \emph{i.e.}, $T(k)\propto (k)^{\alpha}$.
As we know, the degree centrality gives the local influence of nodes, whereas the coreness centrality depicts the global influence of nodes. Hence, we make the following \textit{hypothesis} that high-coreness nodes present more intense attraction than high-degree nodes, and in this case we \textit{hypothesize} that the preferential attachment rate $T(c)$ presents a monotonically \textit{exponential distribution}, \emph{i.e.},
\begin{equation}\label{Tc}
T(c) \propto e^{\beta\cdot c},
\end{equation}
where $\beta$ is the exponent. Similarly, to obtain a smooth curve from noisy data, we take the cumulative function form instead of $T(c)$:
\begin{equation}\label{eq8}
\kappa(c)=\int_0^cT(c)dc.
\end{equation}
According to Eqs. (\ref{Tc}) and (\ref{eq8}), $\kappa(c)$ should be proportional to $e^{\beta\cdot c}$, namely
\begin{equation}\label{kappac}
\kappa(c) \propto e^{\beta\cdot c}.
\end{equation}
Note that, if the exponent $\beta$ is positive, there would be a preferential attachment. 

Shortly, we will use real-world temporal networks to check our hypothesis that the preferential attachment rate $\kappa(c)$ presents an \textit{exponential} distribution: 
\begin{equation}\label{kc}
\kappa(c) \propto e^{\beta\cdot c},    
\end{equation}
where $\beta$ is the exponential exponent.
We will fit the empirical curve from the previous statistical analysis and then compare it against this hypothesized curve of preferential attachment.
\begin{figure*}[t]
\centering
\includegraphics[width=1\textwidth]{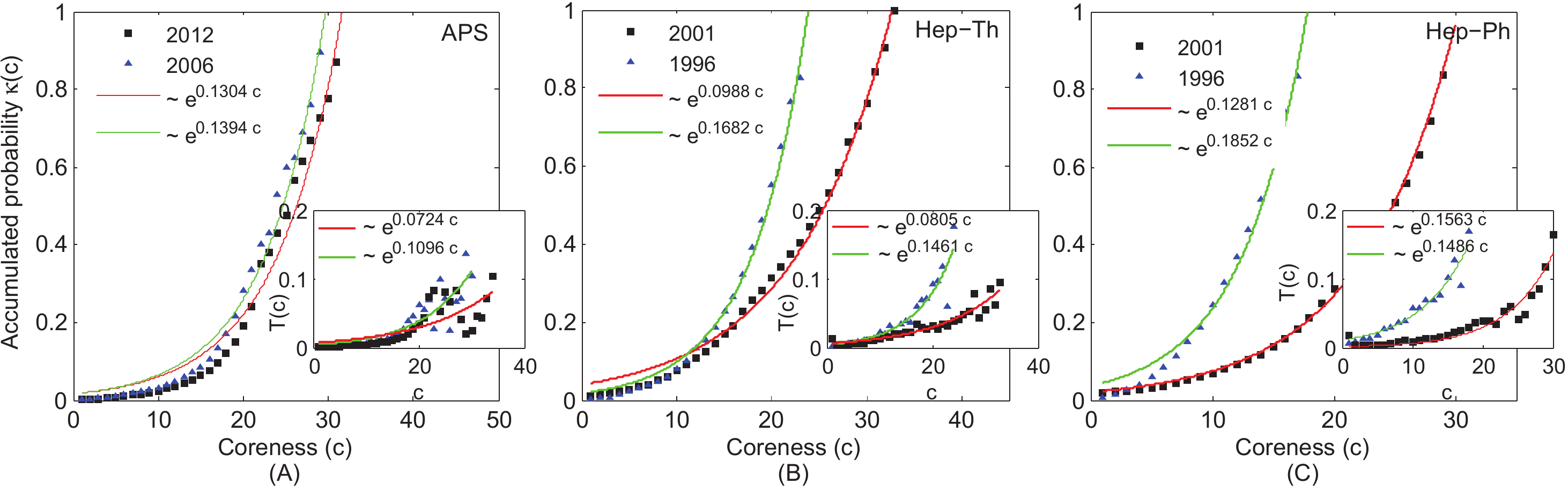}
\caption{\small{The preferential attachment rate $\kappa(c)$ with respect to coreness $c$. Note that, $\kappa(c)$ exhibits an exponential dependence on $c$ on each dataset. Insets: the distribution of the probability $T(c)$ against node coreness $c$. It generally presents to be exponential. For the remaining of this paper, if not mentioned, $\Delta t$ represents one year.}}
\label{preferCore}
\end{figure*}

\subsection{Empirical Analysis}
We perform our analysis on three temporal networks which contain collections of all papers published by three scientific journals: the American Physical Society (APS) Physical Review journals from 1894 to 2013, the High Energy Physics-Phenomenology (Hep-Ph) journal from 1992 to 2002, and the High Energy Physics-Theory (Hep-Th) journal from 1992 to 2003. Each paper is recorded as a data entry including the title, date of publication, and reference(s) to other papers within the dataset. Each entry is denoted as $i$. If paper $i$ cites paper $j$, then there is an edge from $i$ to $j$, denoted as $a_{ij}$ = 1, otherwise $a_{ij}$ = 0. Table I lists the basic statistics of these datasets. In this section, our main concerns are to observe whether new papers would be more likely to attach to core papers (\textit{i.e.}, nodes with high coreness), and explore the dominance of coreness in the preferential attachment process of network growth. 
\begin{table}[b]\label{table1}
\caption{\small{Statistics of the networks used in this paper.}}
\centering
\begin{tabular}{@{\vrule height 10.5pt depth4pt width0pt}lcrcc}
\\
\noalign{\vskip-11pt}
\vrule depth 6pt width 0pt Network & \multicolumn1c{\it N} & \multicolumn1c{\it E} & $\alpha$ & $\beta$ \\
\hline
APS & 527,494 & 5,995,492 & 0.71$\pm$0.31 & 0.30$\pm$0.18\\
Hep-Th & 27,770 & 352,807 & 0.70$\pm$0.12 & 0.19$\pm$0.14\\
Hep-Ph & 30,566 & 347,414 & 0.72$\pm$0.19 & 0.25$\pm$0.17\\
\hline
\end{tabular}\\
\small{Note: $\alpha$ and $\beta$ denote the exponents in the degree-based and coreness-based preferential attachment rates, respectively.}
\end{table}

\subsubsection{Exponential Preferential Attachment Rate}
We first examine our hypothesis of the preferential attachment rate $\kappa(c)$ presents an \textit{exponential distribution} on these temporal networks. Fig. \ref{preferCore} displays the cumulative distribution function $\kappa(c)$ with respect to the coreness $c$ of the nodes that a newly added node attached in the example years. As we can see, $\kappa(c)$ shows an exponential dependence on the coreness $c$ of nodes in each of the datasets, namely $\kappa(c)\propto e^{\beta\cdot c}$. According to our analysis in Section \ref{subsec:PAMCore}, the positive exponents $\beta$ in Fig. \ref{preferCore} indicate that new links are more likely to attach the globally influential nodes, \textit{i.e.}, new links preferentially attach to the nodes with high coreness in networks. To have a clearer look at the exponential dependence of $\kappa(c)$ on $c$, we also plot the distribution of the probabilities $T(c)$ as insets in Fig. \ref{preferCore}. Note that, the probability $T(c)$ also exhibits an exponential dependence on $c$ for each of the datasets. The exponential dependencies of $T(c)$ and $\kappa(c)$ on coreness $c$ exactly verify our conjecture of the exponential preferential attachment to high-coreness nodes in Section \ref{subsec:PAMCore}. The little inconsistence of the exponents between $\kappa(c)$ and $T(c)$ in Fig. \ref{preferCore} is due to the noise of the data. Furthermore, we see that the exponent $\beta$ varies in $[$0.1, 0.2$]$, indicating that the great power of the global influence of nodes in attracting new links. Additionally, it is seen that $\beta$ decreases in the latest years for each of the datasets, which implies that the power of the global influence of nodes decreases gradually in network growth.
\begin{figure}[t]
\centering
\includegraphics[width=0.37\textwidth]{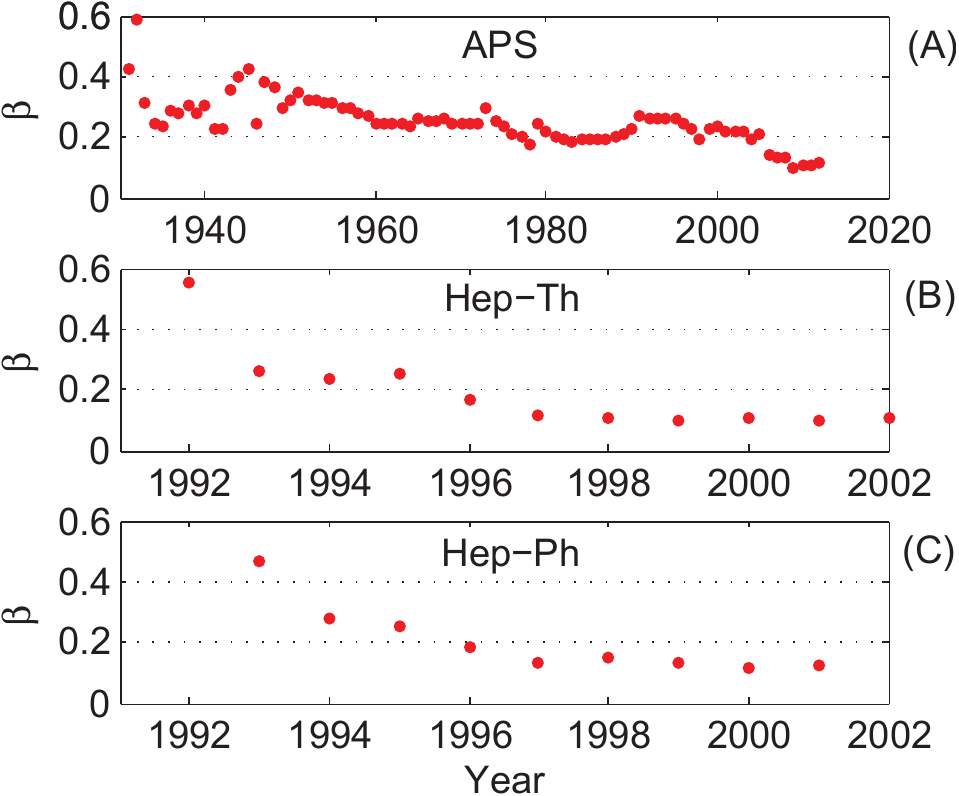}
\caption{\small{The evolution of exponents $\beta$ in the cumulative probability function $\kappa(c)\propto e^{\beta\cdot c}$ over time.}}
\label{beta_accu}
\end{figure}

Fig. \ref{beta_accu} shows the variation of the exponent $\beta$ in the growth of each network over time. Generally, $\beta$ decreases over time and reaches to be stable around 0.12 near the end. The decrease in the exponent $\beta$ indicates that the \textit{intensity} of the preferential attachment to high-coreness nodes gradually becomes less pronounced over time. In other words, new links do preferably attach to globally influential nodes in network growth, but this trend becomes less intensive and reaches a stable state near the end of our observed datasets. Shortly, we will compare coreness-based and degree-based preferential attachments and from the comparison we will see that this phenomenon is caused by the interplay between the local and global influence of nodes. Furthermore, we will see some hidden phenomena happening in the process of network growth. That is, when the network grows bigger, many local cores are formed which relatively have lower coreness but host more connections. 
\begin{figure*}[t]
\centering
\includegraphics[width=1\textwidth]{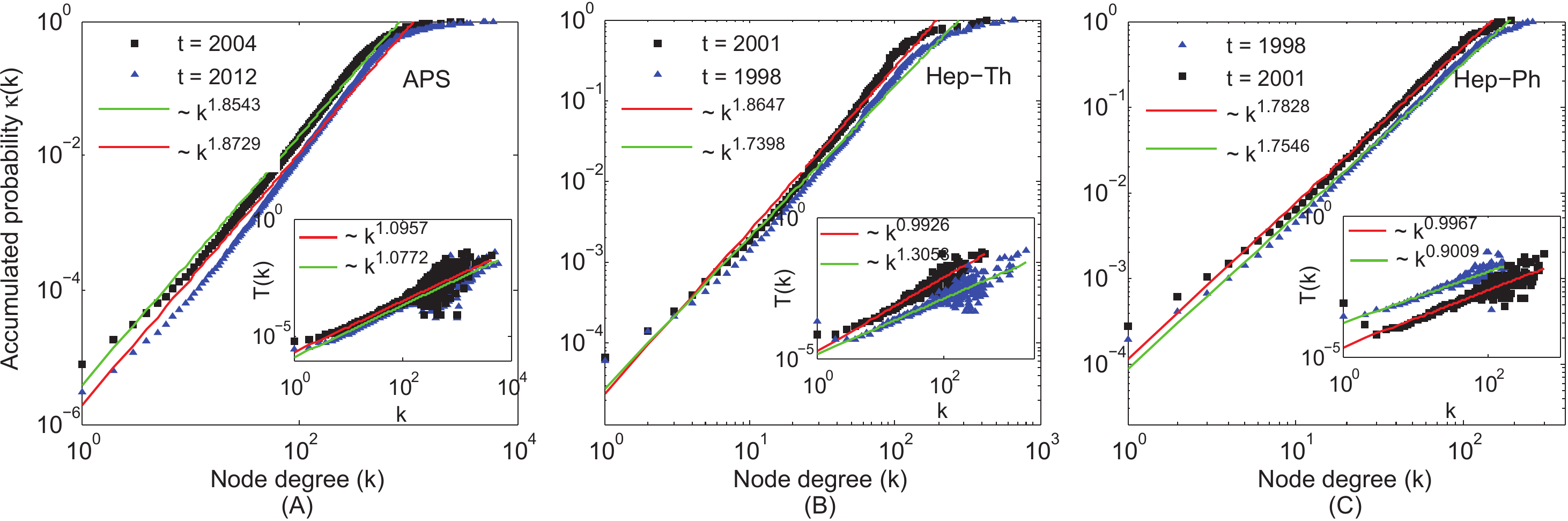}
\caption{\small{The preferential attachment rate $\kappa(k)$ with respect to degree $k$. Note that, $\kappa(k)$ shows power-law dependence on node degree $k$ on each dataset. Insets: the distribution of the probability $T(k)$ against node degree $k$. It generally presents to be power-law on $k$.}}
\label{preferDeg}
\end{figure*}

\subsubsection{Comparison to Degree-based Preferential Attachment}\label{422}
Now, we compare coreness-based preferential attachment to degree-based preferential attachment. Fig. \ref{preferDeg} shows the cumulative probability function $\kappa(k)$ with respect to node degree $k$ for each of the datasets. As we can see, $\kappa(k)$ shows a \textit{power-law dependence} on node degree $k$, namely, $\kappa(k)\propto k^{\alpha+1}$. 
According to Section \ref{growth}, the positive exponents $\alpha$ in Fig. \ref{preferDeg} indicate that nodes with higher degree present greater power in attracting new links. From a mathematical point of view, the increasing rate of exponential functions is significantly faster than that of power-law functions. That is to say, even though the local influence (\textit{i.e.}, degree) also presents to be powerful in attracting new links, the global influence (\textit{i.e.}, coreness) of nodes exhibits to be more powerful compared with the local influence of nodes. This is attributed to the dominance of the global influence over the local influence of nodes in network growth. 
This outcome is also consistent with our expectation for paper citations in scientific journals. Furthermore, we plot the probability distribution $T(k)$ with respect to node degree $k$ as insets in Fig. \ref{preferDeg}. $T(k)$ generally follows a power-law distribution on degree $k$ in each dataset. This result verifies the power-law dependence of the cumulative probability $\kappa(k)$ on node degree $k$ in Eq. (\ref{eq14}). In addition, we can see that the exponent $\alpha$ increases in the latest years for each of the datasets.

\begin{figure}[t]
\centering
\includegraphics[width=0.37\textwidth]{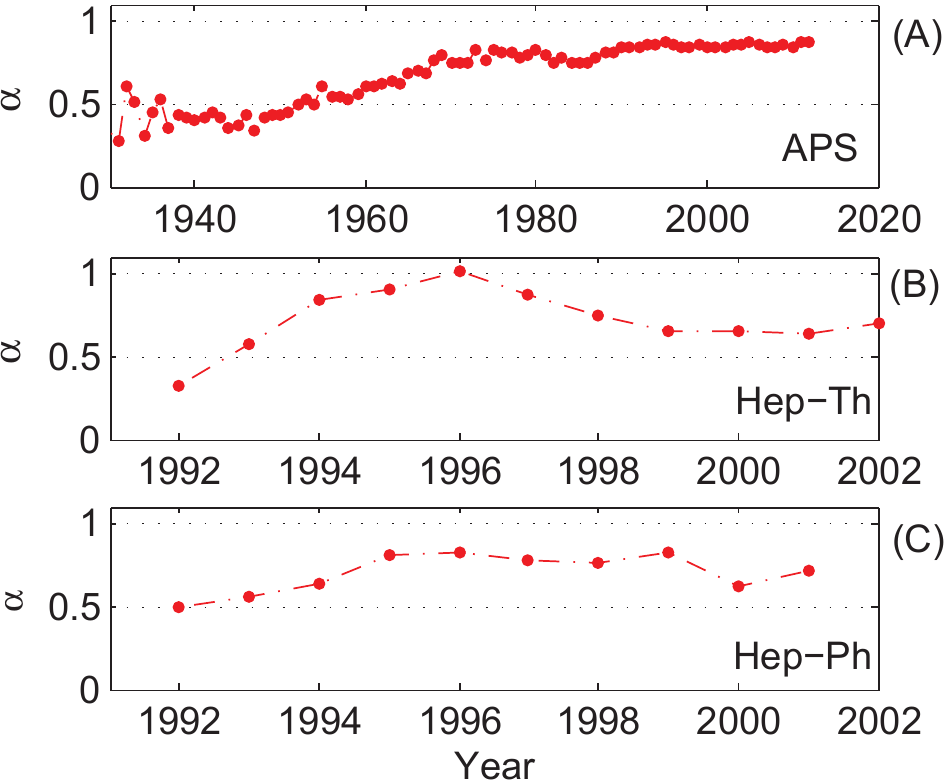}
\caption{\small{The evolution of power-law exponents $\alpha$ in the cumulative probability function $\kappa(k)\propto k^{\alpha}$ over time.}}
\label{alpha_accu}
\end{figure}

Fig. \ref{alpha_accu} shows the variation of the exponent $\alpha$ in network growth for each dataset. We see that, $\alpha$ generally increases from under 0.5 to above 0.7 in spite of some fluctuations, and it reaches to be stable near the end. The increase in exponent $\alpha$ implies the increase in the power of the local influence of nodes in network growth. Combining with the result of the decrease in the exponent $\beta$ in $\kappa(c)\propto e^{\beta\cdot c}$ in Fig. \ref{beta_accu}, these results indicate that the power of node degree in attracting new links increases over time and the influence of coreness decreases. Both $\alpha$ and $\beta$ reach to be stable in network growth, suggesting a state of \textit{equilibrium between the local and global influence of nodes} in network growth.

\subsubsection{Relation between the Degree and Coreness of Nodes}
\begin{figure*}[t]
\centering
\includegraphics[width=1\textwidth]{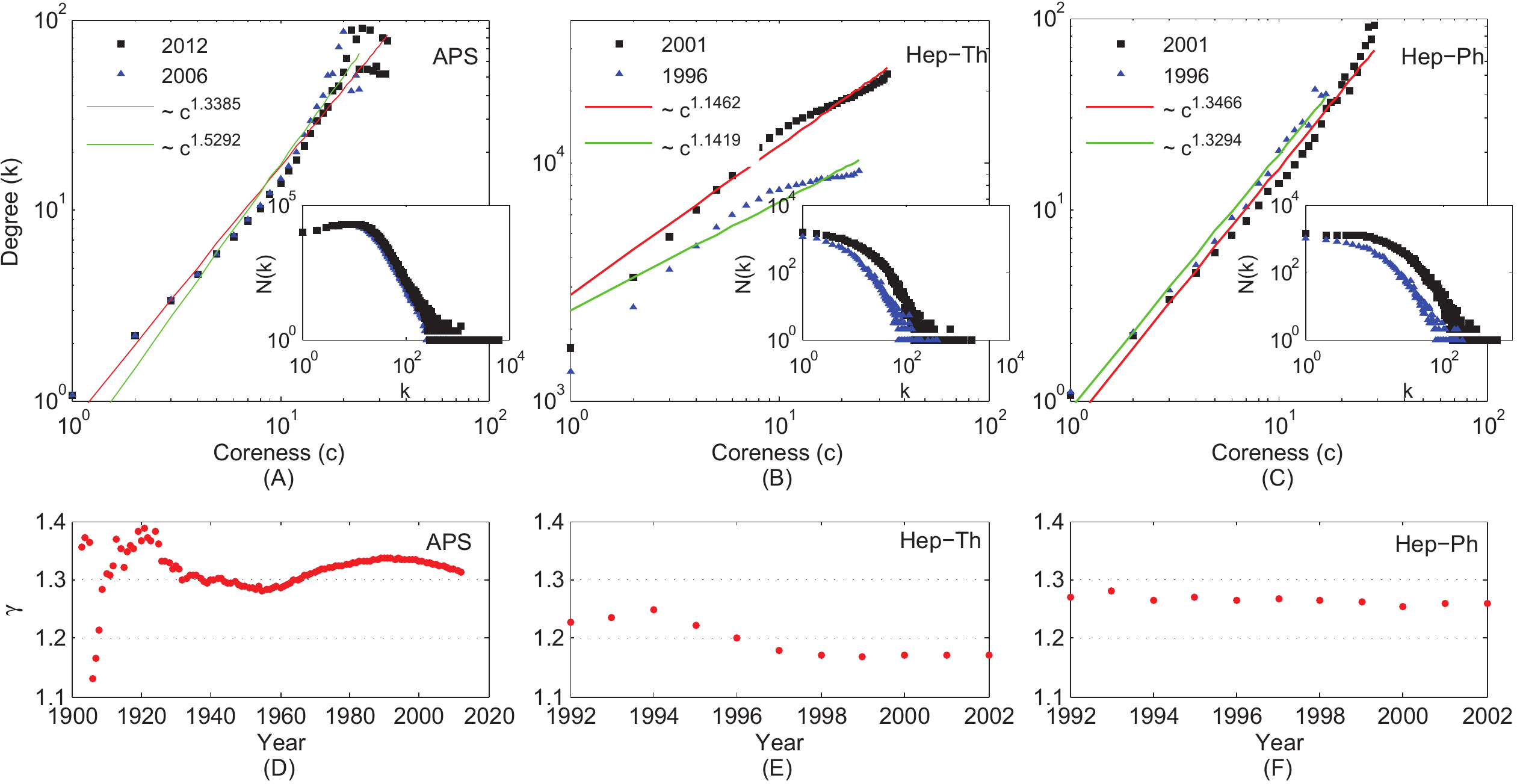}
\caption{\small{(A) - (C): the average degree $\left\langle k_c \right\rangle$ of node on shell $c$. $\left\langle k_c \right\rangle$ follows roughly a power-law dependence on $c$, namely $k_c \propto c^{\gamma}$. Insets: the distribution of node degree in the latest years of each dataset. Note that, when node degree is sufficiently large, the distribution shows to be power-law. (D) - (F): the dynamics of the exponent $\gamma$ in the power-law dependence of $k_c\approx c^{\gamma}$ on $c$ over time. }}
\label{coreVSdegree}
\end{figure*}
In previous analysis, we see that node coreness (wrt. global influence of nodes) exhibits exponential rate in attracting new links and node degree (wrt. local influence of nodes) exhibits power-law rate in attracting new links, and coreness and degree interact with each other over time and reach to a state of equilibrium over time. Now, we move on to take a closer look at the relationship between the coreness and degree of nodes (\textit{i.e.}, the relation between the global and local influence of nodes in network growth), especially their relation in network growth. To obtain a smooth curve from noisy data, we examine \textit{the average degree} $\left\langle k_c \right\rangle$ of nodes on \textit{each shell} $c$. The results on the temporal citation networks in the latest years of the datasets are shown in Fig. \ref{coreVSdegree} (A) - (C). As we can see, the average degree $\left\langle k_c\right\rangle$ follows roughly a power-law dependence on coreness $c$ before $c$ approaches the innermost shells (i.e., large coreness values), namely
\begin{equation}\label{eq13}
\left\langle k_c \right\rangle\propto c^{\gamma},~\gamma>1.
\end{equation}
This equation indicates that, with the linear increase in the global influence of nodes, the local influence averagely increases in a power-law speed. Hence, \textit{the global influence has priority than local influence in attracting new links}. The power-law preferential attachment rate of $\kappa(k)$ on $k$ and the priority of global influence than local influence leads to the intensively increasing rate of $\kappa(c)$ on $c$, i.e., an exponential increasing rate of $\kappa(c)$ on $c$ in Eq. (\ref{kc}). We also plot the distribution of the degrees of nodes in the latest years for each dataset as insets in Fig. \ref{coreVSdegree} (A) - (C). It is seen that when the degree is sufficiently large, the distribution shows to be power-law, which means that the networks tend to be scale-free in each dataset.

We further show in Fig. \ref{coreVSdegree} (D) - (F) the evolution of the power-law exponents $\gamma$ in the relation of $\left\langle k_c \right\rangle \propto c^{\gamma}$ in Eq. (\ref{eq13}). The greater the exponent $\gamma$, the more intensive of global influence prior to local influence. We see that the exponent $\gamma$ fluctuates over time, and it generally shows a descending trend and becomes relatively steady near the end. This implies that the relation between the global and local influence of nodes fluctuates with time, the priority of the global influence becomes less pronounced, and the global influence and local influence gradually reach to a stable level near the end, which also verifies the temporal preferential attachment phenomena in Fig. \ref{beta_accu} and Fig. \ref{alpha_accu}.

\section{Hybrid Preferential Attachment}\label{hybrid}
In this section, we first develop a novel method for measuring the hybrid preferential attachment to both coreness and degree of nodes. Then, we perform empirical analysis on real-world temporal networks. The basic idea is to investigate whether new links are more likely to attach to high-coreness nodes if the nodes are with the same degree, and vise versa.

\subsection{Preferential Attachment Model based on Coreness and Degree}
We use $T(c,k)$ to denote the ``relative'' probability that, within a short period $\Delta t$, a newly added link connects to an existing node with degree $k$ and coreness $c$ before time $t$. Note that, $T(c,k)$ has two variables: the coreness $c$ and degree $k$ of a node. We use $P_{c,k}(t)$ to denote the corresponding time-dependent absolute probability. Note that, $P_{c,k}(t)$ is proportional to $T(c,k)n_{c,k}(t)/N(t)$, where $n_{c,k}(t)$ is the number of nodes with degree $k$ and coreness $c$, and $N(t)$ is the number of nodes immediately before time $t$. $T(c,k)$ can be calculated from the following formulation,
\begin{equation}\label{Tck1}
T(c,k)=\sum_{i,\lambda}^{c_{\lambda}=c,k_{\lambda}(t)=k}a_{i\lambda}\cdot\frac{N(t)}{n_{c,k}(t)},
\end{equation}
where $a_{i\lambda}$ = 1 if node $i$ is newly added and it attaches to node $\lambda$ within the time period $\Delta t$, otherwise 0, and, $c_{\lambda}=c$ and $k_{\lambda}(t)=k$ mean that node $\lambda$ has coreness $c$ and degree $k$ immediately before time $t$. $T(c,k)$ can be normalized as
\begin{equation}\label{Tck}
T(c,k)=\frac{T(c,k)}{\sum_{c,k}T(c,k)}.
\end{equation}
Later, we will derive a general form for $T(c,k)$, in which by integrating $T(c,k)$ with respect to $k$ we will give the form of $T(c)$ and by integrating $T(c,k)$ with respect to $c$ we will give the form of $T(k)$.

From Eq. (\ref{Tck}), we can also examine the localized preferential attachment within each shell of a network. Given a shell, say $c_0$, we can take the cumulative probability function for all nodes, which could have different degrees, in shell $c_0$:
\begin{equation}\label{phi}
\Phi(c_0,k)=\int_0^kT(c_0,k)dk.
\end{equation}
Similarly, when we take a look at the nodes with a fixed degree, say $k_0$, we can examine the localized preferential attachment to these nodes in different shells. We can take the cumulative probability function for different core shells containing the nodes of degree $k_0$:
\begin{equation}\label{pi}
\Pi(c,k_0)=\int_0^cT(c,k_0)dc.
\end{equation}
Later, we will use $\Phi(c_0,k)$ and $\Pi(c,k_0)$ to study the localized preferential attachment within and among shells, respectively, and explore the interplay between the global and local influence of nodes in network growth.

\subsection{The Joint Probability for Hybrid Preferential Attachment}
In this section, we analyze the hybrid preferential attachment with respect to both the coreness and degree of nodes. More specifically, we will derive a general formula for the probability $T(c,k)$ in Section \ref{hybrid}, that a newly added link connecting to an existing node of degree $k$ and coreness $c$ within a short period $\Delta t$. In essence, $T(c,k)$ is a joint probability of $c$ and $k$. Hence, we expect that, integrating $T(c,k)$ with respect to $k$ will produce $T(c)$ in Eq. (\ref{Tc}) and integrating $T(c,k)$ with respect to $c$ will produce $T(k)$ in Eq. (\ref{eq3}). Then, we will study the localized preferential attachment rate $\Phi(c_0,k)$ with respect to the degree $k$ of the nodes on the same shell $c_0$, and the localized preferential attachment rate $\Pi(c,k_0)$ with respect to the coreness $c$ of the nodes having the same degree $k_0$. 

In Section \ref{subsec:PAMCoreness}, we find that the probability of existing nodes attracting new links generally follows an exponential dependence on node coreness, $T(c)\propto e^{\beta c}$, while at the same time, follows a power-law dependence on node degree, $T(k)\propto k^{\alpha}$. Hence, the probability $T(c,k)$ of hybrid preference attachment should be able to produce $T(c)$ by integrating $T(c,k)$ with respect to $k$ and produce $T(k)$ by $T(c,k)$ with respect to $c$. Therefore, we hypothesize that $T(c,k)$ is proportional to $e^{\beta c+\alpha\mbox{ln}(k)}$, namely,
\begin{equation}\label{newTck}
T(c,k) \propto e^{\beta c+\alpha\mbox{ln}(k)},
\end{equation}
where $\alpha$ and $\beta$ are parameters, and $\mbox{ln}(k)$ is the natural logarithm function of $k$. By integrating both sides of Eq. (\ref{newTck}) with respect to $k$, we can get the probability $T(c)$ as follows,
\begin{equation}
T(c) \propto \int_0^k T(c,k) dk \propto \frac{k^{\alpha+1}}{\alpha+1} e^{\beta c}.
\end{equation}
This exactly gives the preferential attachment rate of $T(c)$ in Eq. (\ref{kappac}). Similarly, by integrating both sides of Eq. (\ref{newTck}) with respect to $c$, we can get the probability $T(c)$ as follows,
\begin{equation}
T(k) \propto \int_0^c T(c,k) dc \propto \frac{e^{\beta c}}{\beta} k^{\alpha}.
\end{equation}
This exactly gives the preferential attachment rate of $T(k)$ in Eq. (\ref{eq14}). Therefore, Eq. (\ref{newTck}) is our general form for the probability $T(c,k)$ in hybrid preferential attachment.

\textit{Remark}: In our hypothesis in Eq. (\ref{newTck}), we note that it is equivalent to assume that coreness ($c$) and degree ($k$) of a node are independent, as $T(c,k) \propto T(c)\cdot T(k)$. This helps simplify the problem and analysis. In fact, in many real-world complex networks, degree ($k$) and coreness ($c$) are related to each other, but their relationship is too complicated to be formalized. This can be observed from the process of getting the $k$-core structure of a network (see Fig. \ref{k-core}). This will be discussed in Section \ref{conclusion}. We will further explore the dependence of degree and coreness and also their joint effect on network growth in our future work.

Furthermore, if we fix $c$, say $c_0$, in the joint probability $T(c,k)$ in Eq. (\ref{newTck}), we can get
\begin{equation}\label{TC0K}
T(c_0,k) \propto e^{\beta c_0+\alpha\mbox{ln}k} \propto e^{\beta c_0}\cdot k^{\alpha} \propto k^{\alpha}.
\end{equation}
$T(c_0,k)$ measures the localized preferential attachment with respect to the degree $k$ of the nodes on the same shell $c_0$. The above equation shows that $T(c_0,k)$ follows a power-law dependence on $k$.
Analogously, if we fix $k$, say $k_0$, in the joint probability $T(c,k)$ in Eq. (\ref{newTck}), we can get
\begin{equation}\label{TCK0}
T(c,k_0) \propto e^{\beta c+\alpha\mbox{ln}k_0} \propto k_0^{\alpha}\cdot e^{\beta c} \propto e^{\beta c}.
\end{equation}
$T(c,k_0) $ measures the localized preferential attachment with respect to the coreness $c$ of the nodes having the same degree $k_0$. This equation shows that $T(c,k_0)$ follows an exponential dependence on $c$. In the next two subsections, we will test these two localized preferential attachment phenomena on the temporal networks and verify Eqs. (\ref{TC0K}) and (\ref{TCK0}).

In fact, according to the $k$-core structure of a network, it categories the nodes with similar global influence into the same shell. On the other hand, based on previous analysis, high-degree nodes and high-coreness nodes both show great power in attracting newcomers in network growth. Therefore, it is expected that the attractiveness for the nodes on the same shell is similar, but those of higher degree are more attractive than those of lower degree. Accordingly, we expect an increasing function of $T$ on $k$:
\begin{equation}\label{eq20}
T(c_0,k_1) \leq T(c_0,k_2),~\mbox{if}~k_1\leq k_2,
\end{equation}
where $k_1$ and $k_2$ denote the degrees of an arbitrary pair of nodes on shell $c_0$. Analogously, for the nodes with the same value of degree, we expect that those on inner shells are more attractive than those on peripheries. Accordingly, we expect an increasing function of $T$ on $c$:
\begin{equation}\label{eq21}
T(c_1,k_0) \leq T(c_2,k_0),~\mbox{if}~c_1\leq c_2,
\end{equation}
where $c_1$ and $c_2$ denote the corenesses of an arbitrary pair of shells containing nodes of degree $k_0$. We will test these two expectations in the following two subsections, and verify the relations in Eqs. (\ref{TC0K}) and (\ref{TCK0}).
\begin{figure*}[t]
\centering
\includegraphics[width=1\textwidth]{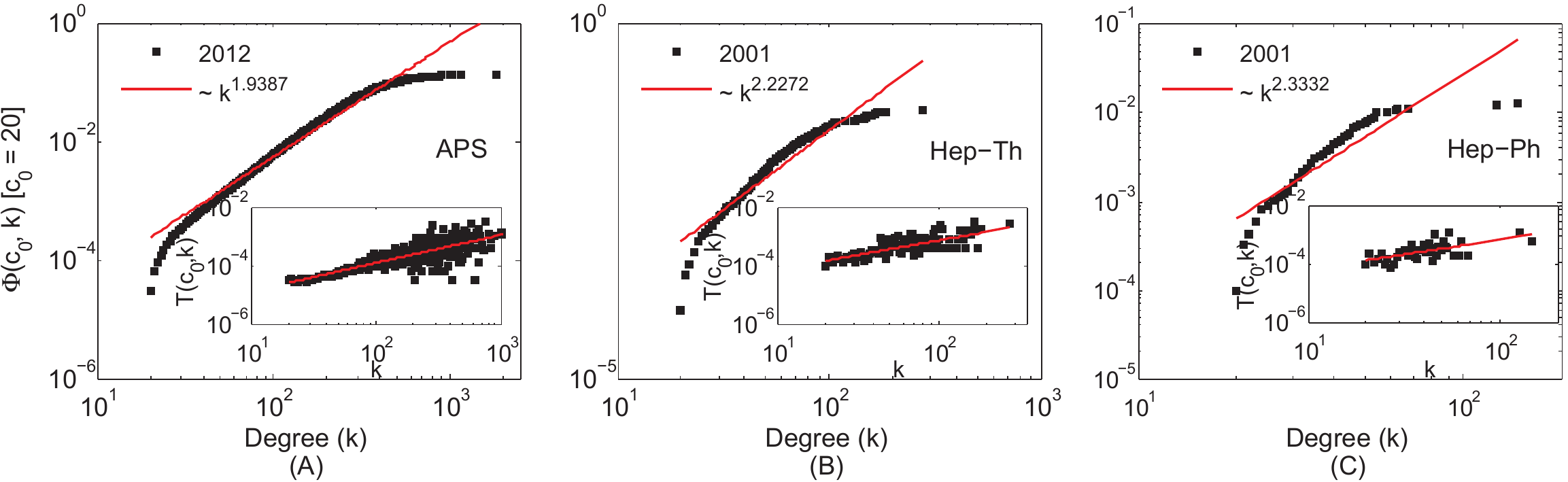}
\caption{\small{Empirical preferential attachment within each shell. Each subfigure shows the preferential attachment rate $\Phi(c_0,k)$ with $c_0$ = 20. Note that, $\Phi(c_0,k)$ shows a power-law dependence on $k$. Insets: the distribution of the probability $T(c_0,k)$ in the latest year of each dataset.}}
\label{coreDeg_within}
\end{figure*}
\begin{figure*}[t]
\centering
\includegraphics[width=1\textwidth]{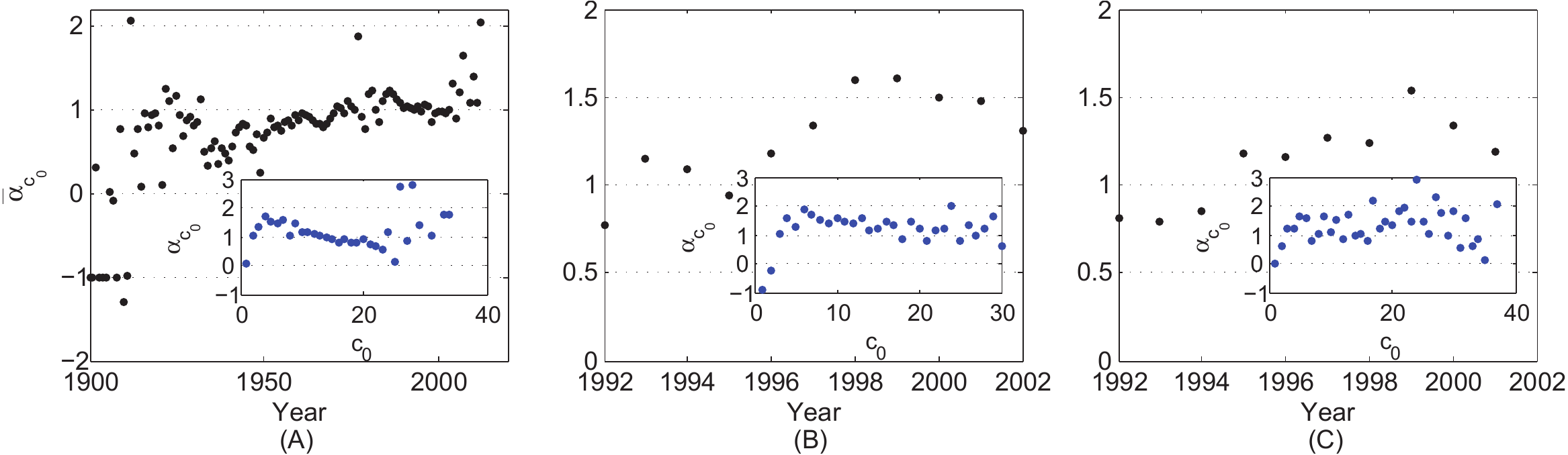}
\caption{\small{The average exponents $\bar{\alpha}_{c_0}$ with respect to the coreness $c_0$ of nodes in the preferential attachment rate $\Phi(c_0,k)\propto k^{\alpha_{c_0}}$ for each year. Overall, $\bar{\alpha}_{c_0}$ presents an increasing trend with $\bar{\alpha}_{c_0}\geq1$ near the end. Insets: the exponents $\alpha_{c_0}$ for each $c_0$ in the latest year.}}
\label{avg_alpha_core_deg}
\end{figure*}

\subsection{Empirical Analysis}\label{SecHybrid}

\subsubsection{Preferential Attachment within Shells}
Here, we first investigate the preferential attachment phenomenon within an arbitrary shell, say $c_0$. In Fig. \ref{coreDeg_within}, we show the cumulative probability function $\Phi(c_0,k)$ when $c_0$ = 20 in the latest year for each of the datasets. As we can see, $\Phi(c_0,k)$ for each dataset does exhibit a power-law dependence on node degree $k$ when $k$ is sufficiently large and before $k$ reaches to the largest value. In order to have a clearer look at the power-law dependence of the cumulative probability function $\Phi(c_0,k)$ on $k$, we also plot the distribution of the probability $T(c_0,k)$ as insets for each dataset in Fig. \ref{coreDeg_within}. Note that $T(c_0,k)$ also presents a power-law dependence on the degree $k$ of the nodes within shell $c_0$ for each dataset:
\begin{equation}
T(c_0,k) \propto k^{\alpha_{c_0}},
\end{equation}
where $\alpha_{c_0}$ is the exponent with respect to $c_0$. The increasing trend of function $T(c_0,k)$ in each dataset verifies our expectation in Eq. (\ref{eq20}), and the power-law dependence of $T(c_0,k)$ on $k$ verifies our hypothesis in Eq. (\ref{TC0K}). According to the integral formula in Eq. (\ref{phi}), the cumulative probability distribution $\Phi(c_0,k)$ should show a power-law dependence on $k$, namely
\begin{equation}
\Phi(c_0,k) \propto k^{\alpha_{c_0}+1}.
\end{equation}
Since $\alpha_{c_0}+1\gg1$ for each dataset in Fig. \ref{coreDeg_within}, we have $\alpha_{c_0}\gg0$. The positive exponents $\alpha_{c_0}$ indicate that, within shell $c_0$, new links preferably attach to the nodes of high degree. That is to say, for the nodes with similar global influence, those with higher local influence are more attractive.

We further show in Fig. \ref{avg_alpha_core_deg} the average exponents with respect to $c_0$ in each year for each of the datasets:
\begin{equation}
\bar{\alpha}_{c_0}=\frac{\sum_{c_0}\alpha_{c_0}}{N(c_0)},
\end{equation}
where $N(c_0)$ denotes the number of all shells. The insets in Fig. \ref{avg_alpha_core_deg}, show the exponent $\alpha_{c_0}$ for each $c_0$ in the latest year of each dataset. As we can see, $\bar{\alpha}_{c_0}$ is generally around 1 or even greater than 1. Compared with the exponents $\alpha$ in $\kappa(k)\propto k^{\alpha}$ in Fig. \ref{alpha_accu}, where $\alpha$ varies in [0.5, 1], $\bar{\alpha}_{c_0}$ is generally greater than $\alpha$, \textit{i.e.},
\begin{equation}
\bar{\alpha}_{c_0}\geq\alpha.
\end{equation}
This inequation implies that, the preferential attachment to the degree of nodes within shells is more intensive than the overall preferential attachment to the degree of nodes. This may be attributed to the overwhelmed power of the global influence of nodes, which results in the local influence being less intensive, but within the same level of global influence, the local influence exhibits its superiority. Additionally, we note in Fig. \ref{avg_alpha_core_deg} that $\bar{\alpha}_{c_0}$ generally displays an increasing trend, which means that the local influence becomes more powerful in attracting new links during the growth of the networks. This is consistent with the results in Fig. \ref{alpha_accu}.
\begin{figure*}[t]
\centering
\includegraphics[width=1\textwidth]{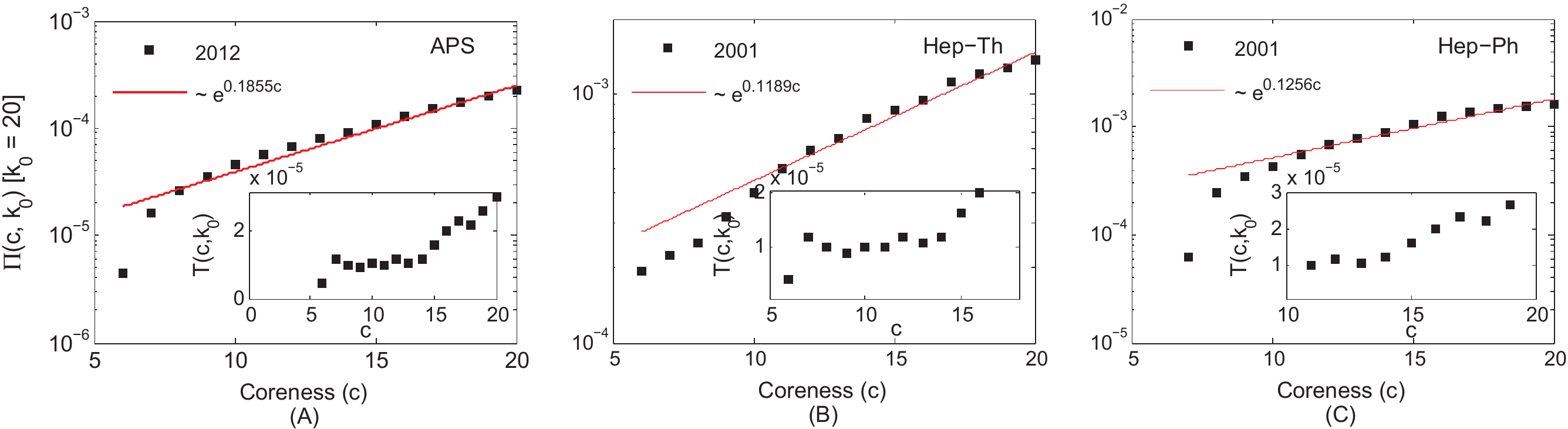}
\caption{\small{Empirical preferential attachment among shells. Each subfigure shows the preferential attachment rate $\Pi(c,k_0)$ with $k_0$ = 20. Note that $\Pi(c,k_0)$ shows an exponential dependence on $c$. Insets: the distribution of the probability $T(c,k_0)$ in the latest year of each dataset.}}
\label{coreDeg_among}
\end{figure*}
\begin{figure*}[t]
\centering
\includegraphics[width=1\textwidth]{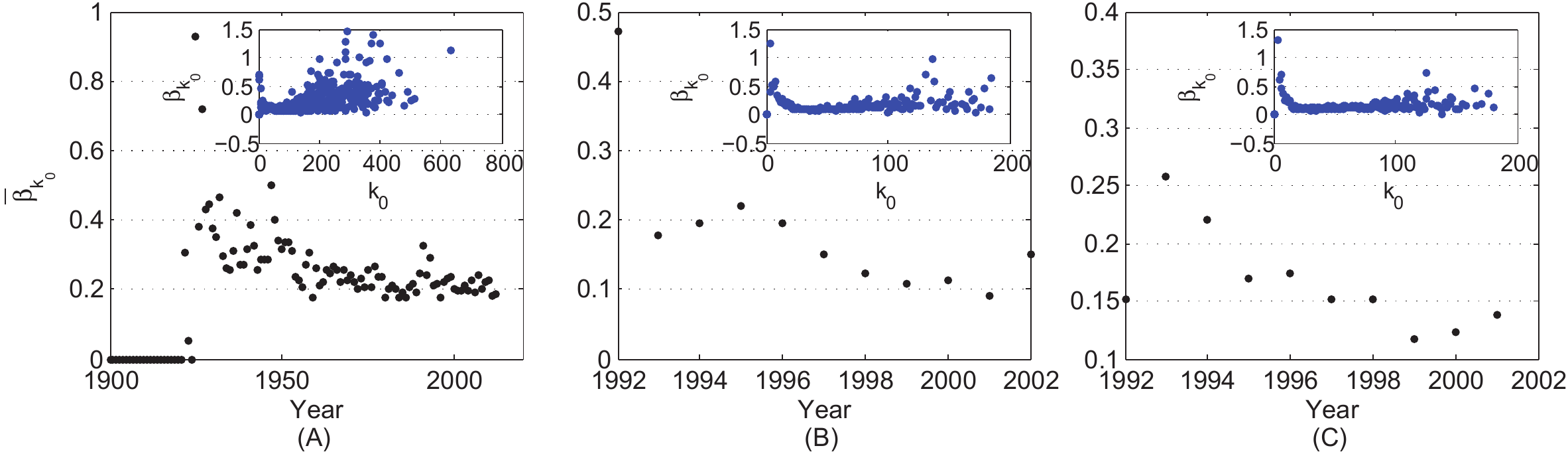}
\caption{\small{The average exponents $\bar{\beta}_{k_0}$ with respect to node degree $k_0$ in the preferential attachment rate $\Pi(c,k_0)\propto e^{c \cdot \beta_{k_0}}$ of each year. Note that, $\bar{\beta}_{k_0}$ generally presents an increasing trend with $\bar{\beta}_{k_0}\geq0.1$. Insets: the exponents $\beta_{k_0}$ for each $k_0$ in the latest year for each dataset.}}
\label{avg_beta_deg_core}
\end{figure*}

\subsubsection{Preferential Attachment among Shells}
Now, we move on to explore the preferential attachment among different shells. More specifically, we study the preferential attachment rate $\Pi(c,k_0)$ for the nodes with the same degree $k_0$ but in different shells. In Fig. \ref{coreDeg_among}, we show the cumulative probability function $\Pi(c,k_0)$ when $k_0$ = 20 in the latest year for each of the datasets. As we can see, $\Pi(c,k_0)$ exhibits an exponential dependence on the coreness $c$ of nodes when $c$ is sufficiently large. In order to have a closer look at the exponential dependence of $\Pi(c,k_0)$ on $c$, we plot in the insets in Fig. \ref{coreDeg_among} the distribution of the probability $T(c,k_0)$ in the latest year of each dataset. Note that $T(c,k_0)$ generally presents an increasing exponential dependence on the coreness $c$ of the nodes who have the same degree $k_0$ for each dataset:
\begin{equation}
T(c,k_0) \propto e^{\beta_{k_0}\cdot c},
\end{equation}
where $\beta_{k_0}$ is the exponent with respect to $k_0$. This increasing trend of function $T(c,k_0)$ in each dataset verifies our expection in Eq. (\ref{eq21}), and the exponential dependence of $T(c,k_0)$ on $c$ verifies our hypothesis in Eq.(\ref{TCK0}). Hence, according to Eq. (\ref{pi}), the cumulative probability $\Pi(c,k_0)$ should present an exponential dependence on $c$, namely
\begin{equation}
\Pi(c,k_0) \propto e^{\beta_{k_0}\cdot c},
\end{equation}
where $\beta_{k_0}$ denotes the exponential exponent. Therefore, the exponential dependence of $\Pi(c,k_0)$ on $c$ in Fig. \ref{coreDeg_among} is expected. The positive exponents $\beta_{k_0}$ in Fig. \ref{coreDeg_among} indicate that, for nodes with the same degree, new links preferably attach to those with higher coreness. 

We further show in Fig. \ref{avg_beta_deg_core} the average exponents with respect to $k_0$ in each year for each of the datasets:
\begin{equation}
\bar{\beta}_{k_0}=\frac{\sum_{k_0}\beta_{k_0}}{N(k_0)},
\end{equation}
where $N(k_0)$ means the number of all different degrees. The insets in Fig. \ref{avg_beta_deg_core} show the exponent $\beta_{k_0}$ for each $k_0$ in the latest year of each dataset. We note that the exponents $\bar{\beta}_{k_0}$ are generally greater than 0.1 and range in [0.1, 0.2]. Compared with the exponents $\beta$ in $\kappa(c)\propto e^{\beta\cdot c}$ in Fig. \ref{beta_accu}, $\bar{\beta}_{k_0}$ is approximate to $\beta$ or slightly lower than $\beta$ near the end:
\begin{equation}
\bar{\beta}_{k_0} \leq \beta.
\end{equation}
This indicates that, the preferential attachment rate $\Pi(c,k_0)$ among shells is similar to the overall preferential attachment rate $\kappa(c)$. This could be attributed to the relatively stable status of the global influence of nodes. The preferential attachment trend with respect to the global influence of nodes keeps stable irrespective of any particular $k_0$. Additionally, we note that $\bar{\beta}_{k_0}$ generally shows a decreasing trend, meaning that the global influence becomes less pronounced in attracting new links over time. This is consistent with the results in Fig. \ref{beta_accu}.

From the analysis in this section, we can conclude that, the attractiveness of a node is not only determined by its global influence in the whole network but also gets impacted by its interaction with local neighbors. Overall, the global influence of nodes tends to be less pronounced over time, which is due to the increase in the local influence of the nodes. This verifies a significant feature in network growth that, when the network grows bigger, many local cores are formed which relatively have lower coreness but host more connections.


\section{Conclusion and Discussion}\label{conclusion}
In this paper, we comprehensively studied the preferential attachment from a new dimension$-$coreness$-$instead of the pure degree-based model. The analysis of this preferential attachment model has been tested in temporal networks. We found that, the probability of existing nodes attracting new links generally follows an exponential dependence on node coreness, which justifies the dominance of node coreness in network growth. Meanwhile, the new dimension discloses some hidden phenomena happening in the process of network growth, that the power of node degree in attracting new links increases over time and the influence of coreness decreases, and finally they reach a state of equilibrium. Furthermore, we derive a general form for the probability of hybrid preferential attachment with respect to both node coreness and degree. This form not only verifies the degree-based and coreness-based preferential attachment rates, but can also guide us in analyzing localized preferential attachment. Our analysis also verifies a significant feature in network growth that, when the network grows bigger, many local cores are formed which relatively have lower coreness but host more connections.

Within this paper, we hypothesize the independency between coreness and degree of network nodes when we derive the general form of hybrid preferential attachment probability $T(c,k)$. However, in many real-world networks, node degree and coreness depends on each other \cite{MK2010}. The dependency between coreness and degree of a node is also an important feature in understanding the network growth. However, due to the dramatic variance in the distributions of node degree and node coreness, the relationship between node degree and coreness is still a big issue in the research field of complex networks. In the future, we will explore the relationship between degree and coreness of nodes through considering different types of networks, such (dis)assortative networks and neutral networks. Then, we will further study the joint effect of node degree and coreness on the growth of complex networks.

\bibliographystyle{IEEEtran}
\bibliography{ref}  

\end{document}